\documentclass[aps,twocolumn,showpacs]{revtex4} 

\newcommand{\stac}[2]{\stackrel{\scriptscriptstyle {#1}}{#2}}

\begin{document}

\title{Baryon number violation due to brane dynamics}

\author{Tomoko Uesugi$^{(1,2)}$, Shunsuke Fujii$^{(3)}$ and Tetsuya Shiromizu $^{(3,4,5)}$}


\affiliation{$^{(1)}$Institute of Physics, Meiji Gakuin University, Yokohama 244-8539, Japan}

\affiliation{$^{(2)}$Department of Physics, Graduate School of Humanities and Sciences, Ochanomizu University, Tokyo 112-8610, Japan}

\affiliation{$^{(3)}$Department of Physics, Tokyo Institute of Technology, Tokyo 152-8551, Japan}

\affiliation{$^{(4)}$Department of Physics, The University of Tokyo,  Tokyo 113-0033, Japan}

\affiliation{$^{(5)}$Advanced Research Institute for Science and Engineering, 
Waseda University, Tokyo 169-8555, Japan}

\date{\today}

\begin{abstract}
We derive the low energy effective theory for two brane system with bulk complex 
scalar fields. Then we find that higher order corrections to the effective equation for 
scalar fields on the brane lead us to the baryon number(B) violating process if the complex scalar 
field is regarded as a particle carrying the baryon number current. We see that the motion of the brane
leads us the baryon number violation via the current-curvature coupling even if the potential for 
scalar fields do not exist. We also estimate the net baryon number assuming 
a C/CP violating interaction. 
\end{abstract}


\maketitle

\label{sec:intro}
\section{Introduction}

According to current various observations, our universe is dominated by matters. We do not 
observe a large amount of anti-matters. 
Since we have no reason for the domination of the matters at the beginning of our universe, 
the origin of this baryon-antibaryon asymmetry is one of important problems in cosmology 
(See Ref. \cite{ReviewB} for the recent review and references). On the other hand, the 
recent progress of superstring theory provides us the new picture of our universe, braneworld. 
Therein our universe is described by the motion of thin wall in higher dimensional spacetimes 
(See Ref. \cite{ReviewBW} for the review). In general the effect of the extra dimension 
is important in the very early universe. 
Since baryogenesis works in the very early universe, 
it is natural to ask if the new type of the realisation of baryogenesis is possible 
in the braneworld context. 
Actually, a couple of ideas have been proposed \cite{SK,DG,DFG}.  

The end of this paper is reexamination of a baryon number violating process discussed in the braneworld 
context \cite{SK}. 
In Ref. \cite{SK}, a scenario of baryogenesis was proposed in Randall-Sundrum type {\rm single} brane 
model \cite{RSII} with 
bulk complex scalar fields. The bulk complex scalar field may be regarded as a particle carrying the 
baryon number current like squarks. Therein the effective action on the brane was derived 
by using the adS/CFT correspondence \cite{GKR}, 
and it was shown that the baryon number violating curvature-current interactions is included 
in the Kaluza-Klein corrections. 
The mechanism is shared with other works in the braneworld \cite{DRT} where 
the escape of massive scalar fields and charge via Kaluza-Klein modes were discussed. 
However, the adS/CFT correspondence leaves us an 
ambiguity in braneworld because it has not been yet established completely and 
``CFT" effective action cannot be determined fully. In this paper, therefore, 
we carefully derive the effective 
equations for Randall-Sundrum two brane system  \cite{RSI} with the bulk complex scalar field 
in order to obtain a definite answer for the effective equation. 
We use the long wave approximation \cite{GE} for the 
derivation. In the case of two branes the adS/CFT correspondence does not hold. Rather say, we can 
solve the bulk field equations under the boundary junction condition without any ambiguities.

The rest of this paper is organized as follows. In Sec. II we describe our models. Then, 
in Sec. III, we solve the bulk equation to derive the effective equation on branes. 
Since we focus 
on baryogenesis, we derive the effective equation for the complex scalar field by using the 
long wave approximation. For simplicity, 
we will assume that contributions from the radion and 
the scalar field to the spacetime dynamics on the brane are negligible. Moreover, we assume that 
the contribution to the correction is dominated by curvature-complex scalar field 
interactions. Using effective equations, in Sec. IV, we will discuss the non-conservation 
of the baryon number current and evaluate the net baryon number to entropy ratio by supposing plausible
CP violating interaction. Finally we give a summary  in Sec. V.

\section{Model}
\label{sec:model}

We consider $Z_2$ symmetric Randall-Sundrum type two brane system with the bulk complex scalar field. 
The bulk and brane actions are given by  
%
\begin{eqnarray}
S_{\rm bulk} & = &  \frac{1}{2\kappa^2} \int d^5x {\sqrt {-g}}({}^{(5)}R-2\Lambda) \nonumber \\
& & -\int  d^5 x {\sqrt {-g}} 
\Biggl[\frac{1}{2}g^{MN}\nabla_M \Phi \nabla_N \Phi^* +V(|\Phi|) \Biggr]
\label{action} 
\end{eqnarray} 
%
and
%
\begin{eqnarray}
S_{\rm brane}^{(\pm)}=\int d^4x {\sqrt {-g_{(\pm)}}} \Bigl[-\sigma_\pm (|\phi_\pm|) +{\cal L}_\pm  \Bigr].  
\end{eqnarray}
%
$g_{MN}$, ${}^{(5)}R$, $\Lambda$ and $\Phi$ are bulk metric, five dimensional Ricci scalar, 
bulk cosmological constant and the U(1) bulk complex scalar field, respectively. 
$g_{(\pm)\mu\nu}$ are the brane induced metric and 
$\phi_\pm$ are complex scalar fields $\Phi$ on each brane. ${\cal L}_\pm $ are the Lagrangian densities 
for matter fields localized on branes. Later we will suppose that the spacetime dynamics on the brane is 
mainly supported by matters and the contribution from the bulk complex scalar field to the 
spacetime dynamics is negligible. 
The bulk stress tensor is 
%
\begin{eqnarray}
T_{MN} & = & \frac{1}{2}(\nabla_M \Phi \nabla_N \Phi^* + \nabla_M \Phi^* \nabla_N \Phi  ) \nonumber \\
& & -g_{MN}\Bigl( \frac{1}{2}|\nabla \Phi|^2 +V(|\Phi|)\Bigr) -\kappa^{-2} \Lambda g_{MN}. 
\end{eqnarray}
%


\section{Effective equation for scalar fields on branes}

In this section, we derive the effective equation for scalar fields on branes using the 
long wave approximation \cite{GE}. We first write down equations for present system and 
then solve them iteratively up to the second order. We would stress that we focus on 
the scalar field $\Phi$ because we are interested in the mechanism of baryogenesis. 

We take the metric ansatz 
%
\begin{eqnarray}
ds^2= e^{2\varphi (x)} dy^2+g_{\mu\nu}(y,x) dx^\mu dx^\nu
\end{eqnarray}
%
and employ (1+4) decomposition. Without a loss of generality, we can suppose that 
branes are located in $y=y_+=0$ and $y=y_-=y_0 >0$. 
Then we expand the metric $g_{\mu\nu}$, the extrinsic curvature 
$K_{\mu\nu}=\frac{1}{2}e^{-\varphi} \partial_y g_{\mu\nu}$ and $\Phi$ in terms of the 
small dimensionless parameter $\epsilon$: 
%
\begin{eqnarray}
& & g_{\mu\nu}= a^2(y,x)(h_{\mu\nu}+\stac{(1)}{g_{\mu\nu}}+\cdots)\\
& & K^\mu_\nu = \stac{(0)}{K^\mu_\nu}+\stac{(1)}{K^\mu_\nu} + \cdots \\
& & \Phi = \stac{(0)}{\Phi}+ \stac{(1)}{\Phi}+\cdots.
\end{eqnarray}
%
Here $\epsilon$ is the square of the ratio of the 
bulk curvature scale $\ell$ to the brane intrinsic curvature scale $L$:
%
\begin{eqnarray}
\epsilon= \Bigl( \frac{\ell}{L}\Bigr)^2.
\end{eqnarray}
%

\subsection{Basic equations}

The ``evolutional" equation is given by 
%
\begin{eqnarray}
e^{-\varphi}\partial_y \tilde K^\mu_\nu & = &  {}^{(4)}\tilde R^\mu_\nu 
-\kappa^2\biggl({}^{(5)}T^\mu_\nu 
-\frac{1}{4} 
\delta^\mu_\nu {}^{(5)}T^\alpha_\alpha \biggr)-K \tilde K^\mu_\nu \nonumber \\
& & - e^{-\varphi}(D^\mu D_\nu e^\varphi )_{\rm traceless}, 
\label{traceless}, 
\end{eqnarray}
%
where $\tilde K^\mu_\nu$ is the traceless part of the extrinsic curvature $K^\mu_\nu$, 
${}^{(4)} \tilde R^\mu_\nu (g)$ is the four dimensional Ricci tensor ${}^{(4)}R^\mu_\nu (g)$, 
and $D_\mu$ is the covariant derivative with respect to $g_{\mu\nu}$. 
$(\cdots)_{\rm traceless}$ stands for the traceless part of $(\cdots)$. 
The equation for the trace part, $K$, is 
%
\begin{eqnarray}
e^{-\varphi}\partial_y K & = &  {}^{(4)}R(g)-\kappa^2 \Bigl(T^\mu_\mu -\frac{4}{3}T^M_M \Bigr)
-K^2 \nonumber \\
& & -e^{-\varphi}D^2 e^\varphi. 
\label{trace}
\end{eqnarray}
%

The constraint equations are 
%
\begin{eqnarray}
& & -\frac{1}{2}\biggl[{}^{(4)}R-\frac{3}{4}K^2+\tilde K^\mu_\nu \tilde K^\nu_\mu \biggr]
=\kappa^2\:{}^{(5)\!}T_{yy} e^{-2\varphi}, 
\label{conK}
\end{eqnarray}
%
and
%
\begin{eqnarray}
D_\nu K^\nu_\mu-D_\mu K = \kappa^2\:{}^{(5)\!}T_{\mu y}e^{-\varphi}. 
\end{eqnarray}
%

The equation for the scalar field is 
%
\begin{eqnarray}
& & e^{-2\varphi} \partial_y^2 \Phi_i +D^\mu  \varphi D_\mu \Phi_i +K e^{-\varphi} \partial_y \Phi_i 
+D^2 \Phi_i \nonumber \\
& & ~~~~-\partial_{\Phi_i} V =0
\end{eqnarray}
%
where $i=1,2$ and $\Phi = \Phi_1 + i \Phi_2$. 
$\Phi_i (i=1,2)$ is the real and the pure imaginary component of the 
complex scalar field $\Phi$. 

By virtue of $Z_2$ symmetry, the junction condition on each brane are given by 
%
\begin{eqnarray}
(K^\mu_\nu - \delta^\mu_\nu K)|_{y=y_\pm} =\frac{\kappa^2}{2}\Bigl(\pm \sigma_\pm \delta^\mu_\nu
\mp \stac{(\pm)}{T^{\mu}_\nu} \Bigr)
\end{eqnarray}
%
and
%
\begin{eqnarray}
e^{-\varphi} \partial_y \Phi_i |_{y=y_\pm}=\pm \frac{1}{2}\tilde \sigma'_\pm (\phi_i^\pm),
\end{eqnarray}
%
where $\phi_i^\pm = \Phi_i(y=y_\pm)$, $\sigma_\pm (\phi_\pm)
= \pm  \sigma_0 +\tilde \sigma_\pm (\phi_\pm) $ and 
$\tilde \sigma'_{\pm i} (\phi_\pm)=\lim_{y \to y_\pm} \partial_{\Phi_i}\tilde \sigma_\pm (\Phi_i)$. 
We set  $\sigma_0$ to be positive. Then the brane at $y=0$($y=y_0$) has the positive(negative) 
tension. The junction condition corresponds to the boundary condition. 

\subsection{Background}

The evolutional equation of $\tilde K^\mu_\nu$ for the background spacetime is 
%
\begin{eqnarray}
e^{-\varphi} 
\partial_y \stac{(0)}{\tilde K^\mu_\nu}=-\stac{(0)}{K}\stac{(0)}{\tilde K^\mu_\nu}. 
\end{eqnarray}
%
The constraint equations are 
%
\begin{eqnarray}
-\frac{1}{2}\Biggl[ -\frac{3}{4} \stac{(0)}{K^2}
+ \stac{(0)}{\tilde K^\mu_\nu}  \stac{(0)}{\tilde K^\nu_\mu}    \Biggr] = \kappa^2 
\stac{(0)}{T}_{yy} e^{-2\varphi}
\end{eqnarray}
%
and
%
\begin{eqnarray}
D_\nu \stac{(0)}{K^\nu_\mu} -D_\mu \stac{(0)}{K}=0. 
\end{eqnarray}
%
The equation for the scalar field becomes
%
\begin{eqnarray}
e^{-2\varphi} \partial_y^2 \stac{(0)}{\Phi}_i +\stac{(0)}{K}e^{-\varphi}\partial_y 
\stac{(0)}{\Phi}_i =0.
\end{eqnarray}
%
The junction conditions are 
%
\begin{eqnarray}
(\stac{(0)}{K}_{\mu\nu}-g_{\mu\nu} \stac{(0)}{K})_{y=y_{(\pm)}} 
= \pm \frac{\kappa^2}{2}\sigma_0 g_{\mu\nu}
\end{eqnarray}
%
and
%
\begin{eqnarray}
\partial_y \stac{(0)}{\Phi_i}|_{y=y_\pm}=0. 
\end{eqnarray}
%
After all, we can easily find the solution for the extrinsic curvature as 
%
\begin{eqnarray}
\stac{(0)}{K^\mu_\nu} = -\frac{1}{\ell}\delta^\mu_\nu 
\end{eqnarray}
%
where $\ell^{-1} = \frac{1}{6}\kappa^2 \sigma_0$. Then the background spacetime is determined 
as  
%
\begin{eqnarray}
\stac{(0)}{g_{\mu\nu}}=a^2(x,y)h_{\mu\nu}(x)
\end{eqnarray}
%
where 
%
\begin{eqnarray}
a(x,y)=e^{-\frac{d(x,y)}{\ell}}
\end{eqnarray}
%
and $d= \int^y_0dy e^{\varphi (x)} = y e^{\varphi (x)}$. 
$h_{\mu\nu} (x)$ is the induced metric on the positive tension brane at $y=0$. 
The zero-th order solution of scalar fields is 
a function only of $x$, 
%
\begin{eqnarray}
\stac{(0)}{\Phi_i}(y,x)= \eta_i(x). 
\end{eqnarray}
%

\subsection{First order}

The evolutional equation of $\tilde K^\mu_\nu$ in the first order is 
%
\begin{eqnarray}
e^{-\varphi} \partial_y \stac{(1)}{{\tilde K}^\mu_\nu}
& = & -\kappa^2 \Bigl(T^\mu_\nu -\frac{1}{4}\delta^\mu_\nu T \Bigr)^{(1)}
-\stac{(0)}{K}  \stac{(1)}{{\tilde K}^\mu_\nu} +({}^{(4)}\tilde R^\mu_\nu)^{(1)} \nonumber \\
& & -\frac{1}{a^2}(e^{-\varphi}{\cal D}^\mu {\cal D}_\nu e^\varphi )_{\rm traceless},
\end{eqnarray}
%
where ${\cal D}_\mu$ is the covariant derivative with respect to the metric $h_{\mu\nu}$. 
$( \cdots )^{(1)}$ stands for the first order part of $(\cdots)$. 
Then the solution is given by 
%
\begin{eqnarray}
\stac{(1)}{{\tilde K}^\mu_\nu}(y,x) & = & -\frac{\ell}{2}a^{-2} {}^{(4)}\tilde R^\mu_\nu (h) \nonumber \\
& & -a^{-2}\Bigl({\cal D}^\mu {\cal D}_\nu d +\frac{1}{\ell}{\cal D}^\mu d{\cal D}_\nu d \Bigr)_{\rm traceless}
\nonumber \\
& & +\frac{\kappa^2 \ell}{4}a^{-2}({\cal D}^\mu \eta {\cal D}_\nu \eta^* +{\cal D}^\mu \eta^* {\cal D}_\nu 
\eta)_{\rm traceless} \nonumber \\
& & +\chi^\mu_\nu (x)a^{-4},
\end{eqnarray}
%
where ${}^{(4)}R_{\mu\nu}(h)$ is the Ricci scalar of the metric $h_{\mu\nu}$, 
$\chi^\mu_\nu (x)$ is the ``constant" of integration, and $\eta = \eta_1+ i \eta_2$. 
$\eta_1$ and $\eta_2$ are real and pure imaginary parts of $\eta$. 
The trace part of the extrinsic curvature 
is directly computed from the Hamiltonian constraint and the result is 
%
\begin{eqnarray}
\stac{(1)}{K} (y,x) & = &  
-\frac{\ell}{6a^2}{}^{(4)}R(h)-\frac{1}{a^2} \Bigl({\cal D}^2 d -\frac{1}{\ell} ({\cal D} d)^2 \Bigr) 
\nonumber \\
& & +\frac{\kappa^2 \ell}{3}\frac{1}{a^2} \Bigl( \frac{1}{2}|{\cal D} \eta |^2+a^2 V(\eta_i) \Bigr).
\end{eqnarray}
%
We can summarise the result as 
%
\begin{eqnarray}
\stac{(1)}{K^\mu_\nu} -\delta^\mu_\nu \stac{(1)}{K}&  = & 
-\frac{\ell}{2}a^{-2}{}^{(4)}G^\mu_\nu (h) -a^{-2}
\Bigl[{\cal D}^\mu {\cal D}_\nu d -\delta^\mu_\nu {\cal D}^2 d \nonumber \\
& & +\frac{1}{\ell} \Bigl( {\cal D}^\mu d {\cal D}_\nu d +\frac{1}{2}\delta^\mu_\nu ({\cal D} d)^2 \Bigr) \Bigr] 
\nonumber \\
& & +\frac{\kappa^2 \ell}{4}a^{-2}
\Bigl[{\cal D}^\mu \eta {\cal D}_\nu \eta^* + {\cal D}^\mu \eta^* {\cal D}_\nu \eta
\nonumber \\
& & -\delta^\mu_\nu (|{\cal D} \eta|^2 +a^2 V(\eta)) \Bigr]
+\chi^\mu_\nu (x) a^{-4}.
\end{eqnarray}
%

Applying the junction condition on the positive tension brane to the above equation, we obtain 
%
\begin{eqnarray}
0 & = & - \frac{\kappa^2}{2}(\tilde \sigma_+ \delta^\mu_\nu -\stac{(+)}{T^{\mu}_\nu})
-\frac{\ell}{2} {}^{(4)}G^\mu_\nu (h) \nonumber \\
& & +\frac{\kappa^2 \ell}{4}
\Bigl[{\cal D}^\mu \eta {\cal D}_\nu \eta^* + {\cal D}^\mu \eta^* {\cal D}_\nu \eta 
-\delta^\mu_\nu (|{\cal D} \eta|^2 +V(\eta)) \Bigr] \nonumber \\
& & +\chi^\mu_\nu (x). 
\end{eqnarray}
%
From the junction condition on the negative tension brane, we also have 
%
\begin{eqnarray}
0 & = & \frac{\kappa^2}{2}(\tilde \sigma_- \delta^\mu_\nu -a_0^{-2}T^{\mu (-)}_\nu)
 -\frac{\ell}{2}a^{-2}_0{}^{(4)}G^\mu_\nu (h) \nonumber \\
& & -a^{-2}_0
\Bigl[{\cal D}^\mu {\cal D}_\nu d_0 -\delta^\mu_\nu {\cal D}^2 d_0 \nonumber \\
& & +\frac{1}{\ell} \Bigl( {\cal D}^\mu d_0 {\cal D}_\nu d_0 +\frac{1}{2}\delta^\mu_\nu ({\cal D} d_0)^2 \Bigr) \Bigr] 
\nonumber \\
& & +\frac{\kappa^2 \ell}{4}a^{-2}_0
\Bigl[{\cal D}^\mu \eta {\cal D}_\nu \eta^* +{\cal D}^\mu \eta^* {\cal D}_\nu \eta \nonumber \\
& & -\delta^\mu_\nu (|{\cal D} \eta|^2 +a_0^2 V(\eta)) \Bigr] 
+\chi^\mu_\nu (x) a^{-4}_0.
\end{eqnarray}
%
Then, eliminating the constant of integration $\chi_{\mu\nu}(x)$ from the above two equations, we 
can obtain the effective gravitational equation on the positive tension brane
%
\begin{eqnarray}
& & (1-a_0^2) {}^{(4)}G_{\mu\nu}(h) \nonumber \\
& & ~~~~=   \frac{2}{\ell}a_0^2 \Bigl[{\cal D}^\mu {\cal D}_\nu d_0 -\delta^\mu_\nu {\cal D}^2 d_0 
+\frac{1}{\ell} \Bigl( {\cal D}^\mu d_0 {\cal D}_\nu d_0 \nonumber \\
& & ~~~~~~+\frac{1}{2}\delta^\mu_\nu ({\cal D} d_0)^2 \Bigr) \Bigr] 
+\kappa^2 \ell^{-1} \Bigl(T^{(+)}_{\mu\nu}+a_0^2 T_{\mu\nu}^{(-)}  \Bigr)
\nonumber \\
& & ~~~~~~+\frac{\kappa^2}{2}(1-a_0^2)\Bigl[{\cal D}^\mu \eta {\cal D}_\nu \eta^* 
+ {\cal D}^\mu \eta^* {\cal D}_\nu \eta -\delta^\mu_\nu (|{\cal D} \eta|^2 \nonumber \\
& & ~~~~~~+(1+a_0^2) V(\eta)) \Bigr] 
-\frac{\kappa^2}{\ell}(\tilde \sigma^++a_0^4 \tilde \sigma^-) h_{\mu\nu}.
\end{eqnarray}
%
This is not our main conclusion. What we want to do in this section 
is the derivation of the effective equation for scalar fields. 

Since we are interested in the interaction between the curvature and scalar fields, 
we omit contributions of the radion field, that is, the derivative of 
$\varphi$, $D_\mu \varphi$. Then the bulk equation in the first order is 
%
\begin{eqnarray}
e^{-2\varphi} \partial_y^2 \stac{(1)}{\Phi_i}+\stac{(0)}{K}e^{-\varphi} \partial_y \stac{(1)}{\Phi_i}
+(D^2 \Phi_i)^{(1)}-\partial_{\eta_i}V(\eta)=0.
\end{eqnarray}
%
By performing the integration over $y$, the above equation becomes 
%
\begin{eqnarray}
\partial_y \stac{(1)}{\Phi_i} = \frac{\ell}{2} \Bigl(a^{-2}{\cal D}^2 \eta_i-\frac{1}{2}
\partial_{\eta_i}V(\eta)  \Bigr) e^\varphi +\frac{\chi_i(x)}{a^4},
\end{eqnarray}
%
where $\chi_i (x)$ is the integral constant. We assume $\stac{(1)}{\Phi_i}(0,x)=0$. 
Applying the junction condition for  
scalar fields to the above equation, we obtain two equations 
%
\begin{eqnarray}
\frac{1}{2}\tilde \sigma'_{+ i} = \frac{\ell}{2} \Bigl({\cal D}^2 \eta_i-\frac{1}{2}
\partial_{\eta_i}V(\eta)  \Bigr) + \chi_i(x) e^{-\varphi}
\end{eqnarray}
%
and
%
\begin{eqnarray}
-\frac{1}{2}\tilde \sigma'_{- i} = \frac{\ell}{2} \Bigl( a_0^{-2}{\cal D}^2 \eta_i-\frac{1}{2}
\partial_{\eta_i}V(\eta)  \Bigr) + \chi_i(x) e^{-\varphi}a_0^{-4}. 
\end{eqnarray}
%
Then, eliminating the integral constant $\chi_i (x)$ from the above two equations, we obtain the 
effective equation for scalar fields in the first order 
%
\begin{eqnarray}
(1-a_0^2){\cal D}^2 \eta_i -\frac{1}{2}(1-a_0^4) \partial_{\eta_i} V(\eta)=\frac{1}{\ell}(\tilde \sigma'_{+ i} 
+a_0^4 \tilde \sigma'_{- i}). 
\end{eqnarray}
%
It is easy to see that the baryon number current is conserved up to this first order. 
Therefore we will consider the second order corrections into the effective equation for  
scalar fields.

\subsection{Second order}

From now on we assume that the contribution from $\eta_i$ to the effective theory 
is negligible compared 
to that from ${}^{(4)}R(D \eta)^2$ and $\stac{(+)}{T}_{\mu\nu}(D \eta)^2$ in the 
calculation of the second order. This is because we are interested in the 
interaction between scalar fields and the brane intrinsic curvature. 

The equation for scalar fields in the second order is 
%
\begin{eqnarray}
& & e^{-2\varphi} \partial_y^2 \stac{(2)}{\Phi_i}+\stac{(0)}{K}e^{-\varphi} \partial_y  \stac{(2)}{\Phi_i}
+\stac{(1)}{K} e^{-\varphi} \partial_y \stac{(1)}{\Phi_i} \nonumber \\
& & ~~~~~~~+(D^2\Phi_i )^{(2)}=0. \label{2ndeqphi}
\end{eqnarray}
%
Then the formal integration of Eq. (\ref{2ndeqphi}) over $y$ gives us 
%
\begin{eqnarray}
\partial_y \stac{(2)}{\Phi_i} & = &  -a^{-4} \int^y dy e^{2\varphi} a^4 
\Bigl( \stac{(1)}{K}e^{-\varphi} \partial_y \stac{(1)}{\Phi_i}+ (D^2\Phi_i )^{(2)}  \Bigr) \nonumber \\
& & +a^{-4} \stac{(2)}{\chi_i}(x). \label{formal}
\end{eqnarray}
%
To proceed the calculation further, we need to determine $\stac{(1)}{g_{\mu\nu}}$ in 
%
\begin{eqnarray}
g_{\mu\nu}=a^2(h_{\mu\nu}+\stac{(1)}{g}_{\mu\nu}).
\end{eqnarray}
%
Assuming $\stac{(1)}{g}_{\mu\nu}=0$ at $y=y_+=0$ and using the results in the 
previous section, we obtain 
%
\begin{eqnarray}
\stac{(1)}{g}_{\mu\nu} & = & -\frac{\ell^2}{2}(a^{-2}-1)\Bigl({}^{(4)}R_{\mu\nu}(h) 
-\frac{1}{6}h_{\mu\nu}{}^{(4)}R(h) \Bigr) \nonumber \\
& & +\frac{\kappa^2}{4}\ell^2 (a^{-2}-1) \Bigl( 
{\cal D}_\mu \eta {\cal D}_\nu \eta^*+{\cal D}_\mu \eta^* {\cal D}_\nu \eta \nonumber \\
& & -\frac{1}{3}h_{\mu\nu}|{\cal D}\eta|^2\Bigr) 
+\frac{\kappa^2}{6}\ell V(\eta) h_{\mu\nu} d \nonumber \\
& & +\frac{\ell}{2}(a^{-4}-1) \chi_{\mu\nu}(x),
\end{eqnarray}
%
%
%
where
%
\begin{eqnarray}
\chi_{\mu\nu}(x) & = &  \frac{\ell}{2}({}^{(4)}G_{\mu\nu}(h)-\kappa^2 \ell^{-1}T_{\mu\nu}^{(+)})
+\frac{\kappa^2}{2}\tilde \sigma_+ (\phi_+) h_{\mu\nu} \nonumber \\
& & -\frac{\kappa^2 \ell}{4} \Bigl[ 
{\cal D}_\mu \eta {\cal D}\eta^*+{\cal D}_\mu \eta^* {\cal D}\eta \nonumber \\
& & -h_{\mu\nu}\Bigl( ({\cal D}\eta)^2+V(\eta) \Bigr) \Bigr].
\end{eqnarray}
%
Then we can calculate $(D^2\Phi_i )^{(2)}$ as 
%
%
%
\begin{eqnarray}
(D^2\Phi_i )^{(2)}
& = &  a^{-2}{\cal D}^2 \stac{(1)}{\Phi_i}- a^{-4}\frac{h^{\alpha\beta}}{2} \stac{(1)}{g}_{\alpha\beta}
{\cal D}_\mu (a^2 {\cal D}^\mu \stac{(0)}{\Phi_i}) \nonumber \\
& & +a^{-4} {\cal D}_\mu \Bigl[a^2(-\stac{(1)}{g^{\mu\nu}}+\frac{h^{\alpha\beta}}{2} 
\stac{(1)}{g}_{\alpha\beta} h^{\mu\nu}  ) {\cal D}_\nu \stac{(0)}{\Phi_i} \Bigr] \nonumber \\
& = & a^{-2}{\cal D}^2 \stac{(1)}{\Phi_i}+\frac{\ell^2}{12}a^{-2}(a^{-2}-1){}^{(4)}R(h){\cal D}^2\eta_i \nonumber \\
& & +a^{-2}{\cal D}_\mu \Biggl[ 
\Biggl\lbrace 
\frac{\ell^2}{2}(a^{-2}-1) \Bigl({}^{(4)}R^{\mu\nu}(h) \nonumber \\
& & -\frac{1}{3}h^{\mu\nu}{}^{(4)}R(h) \Bigr) \nonumber \\
& &  -\frac{\ell}{2}(a^{-4}-1)\chi^{\mu\nu}
\Biggr\rbrace {\cal D}_\nu \eta_i
\Biggr]
\end{eqnarray}
%
%
%
Substituting this for Eq. (\ref{formal}), we see 
%
\begin{eqnarray}
e^{-\varphi} \partial_y \stac{(2)}{\Phi_i} & = & 
-{\cal D}^2 \Biggl[ \frac{\ell^2}{16} (a^{-6}+a^{-2})\tilde \sigma'_{+i} \nonumber \\
& & +\frac{\ell^3}{4} \Biggl\lbrace 
a^{-4} \Bigl(\frac{d}{\ell}+\frac{1}{2} a^2 \Bigr) -\frac{1}{4}(a^{-6}+a^{-2})
\Biggr\rbrace {\cal D}^2 \eta_i \nonumber \\
& & +\frac{\ell^3}{8}\Biggl\lbrace 
a^{-2}\Bigl(\frac{d}{\ell}+\frac{1}{2} \Bigr)+\frac{1}{4}(a^{-6}+a^{-2}) \Bigr)
\Biggr\rbrace \partial_{\eta_i}V (\eta) \Biggr] \nonumber \\
& & 
+\frac{\ell^3}{48}a^{-2} {}^{(4)}R(h) \partial_{\eta_i}V(\eta) \nonumber \\
& & +\frac{\ell^2}{12}a^{-6} \chi_i e^{-\varphi} {}^{(4)}R(h) -\frac{\ell^3}{24} a^{-2}{}^{(4)}R {\cal D}^2 \eta_i \nonumber \\
& & + {\cal D}_\mu \Biggr[ \Biggl\lbrace 
-\frac{\ell^3}{2}\Bigl(\frac{d}{\ell}a^{-4}+\frac{1}{2}a^{-2} \Bigr)
\Bigl( {}^{(4)}R^{\mu\nu}(h) \nonumber \\
& & -\frac{1}{3}h^{\mu\nu}{}^{(4)}R(h) \Bigr) 
+\frac{\ell^2}{4}(a^{-6}+a^{-2}) \chi^{\mu\nu} \Biggr\rbrace {\cal D}_\nu \eta_i 
\Biggr] \nonumber \\
& & +\stac{(2)}{\chi_i} a^{-4} e^{-\varphi}.
\end{eqnarray}
%
In the above we used the solution to $\stac{(1)}{\Phi_i}$
%
\begin{eqnarray}
\stac{(1)}{\Phi_i}(y,x)&  = &  \frac{\ell}{8}(a^{-4}-1)\tilde \sigma'_{+i} \nonumber \\
& & +\frac{\ell^2}{4}\Bigl[(a^{-2}-1)-\frac{1}{2}(a^{-4}-1)  \Bigr]{\cal D}^2 \eta_i \nonumber \\
& & -\frac{\ell^2}{4} \Bigl[\frac{d}{\ell}-\frac{1}{4}(a^{-4}-1) \Bigr] \partial_{\eta_i}V.
\end{eqnarray}
%
Using the junction condition and following the same argument as the previous section, 
we obtain the effective equation for the scalar field 
up to the second order, 
%
\begin{eqnarray}
& & (1-a_0^2) {\cal D}^2 \eta_i -\frac{1}{2}(1-a_0^4) \partial_{\eta_i}V(\eta_i) \nonumber \\
& & ~~~=  \ell^{-1} (\tilde \sigma'_{+i}+a_0^4\tilde \sigma'_{-i} )
-\frac{\ell^2}{24}(1-a_0^2){}^{(4)}R(h) \partial_{\eta_i}V(\eta) \nonumber \\
& & ~~~~~-\frac{\ell}{6}(1-a_0^{-2})\chi_i e^{-\varphi} {}^{(4)}R(h) 
+\frac{\ell^2}{12}(1-a_0^2)
{}^{(4)}R(h){\cal D}^2 \eta_i 
\nonumber \\
& & ~~~~~+ {\cal D}_\mu 
\Biggl[ \Biggl\lbrace
\ell^2 \Bigl(-\frac{d_0}{\ell}+\frac{1}{2}(1-a_0^2)  \Bigr)
\Bigl(
{}^{(4)}R^{\mu\nu}(h) \nonumber \\
& & ~~~~~-\frac{1}{3}h^{\mu\nu}{}^{(4)}R(h) \Bigr) 
-\frac{\ell}{2}(2-a_0^{-2}-a_0^2) \chi^{\mu\nu} \Biggr\rbrace {\cal D}_\nu \eta_i
\Biggr] \nonumber \\
& & ~~~~~+ {\cal D}^2\Biggl[ 
\frac{\ell}{8}(2-a_0^{-2}-a_0^2) \tilde \sigma'_{+i}
+\frac{\ell^2}{2}\Biggl\lbrace 
-\frac{d_0}{\ell}+\frac{1}{2}(1-a_0^2) \nonumber \\
& & ~~~~~-\frac{1}{4}(2-a_0^{-2}-a_0^2)
\Biggr\rbrace {\cal D}^2 \eta_i
+\frac{\ell^2}{4}
\Biggl\lbrace
-\frac{d_0}{\ell}a_0^2 \nonumber \\
& & ~~~~~+\frac{1}{2}(1-a_0^2)+\frac{1}{4}(2-a_0^{-2}-a_0^2)
\Biggr\rbrace \partial_{\eta_i}V
\Biggr]
\end{eqnarray}
%
%
%
Now we can see the coupling between the brane intrinsic curvature and the scalar field. 
As seen in the next section, 
this gives us the possibility of the baryon number violation due to the spacetime dynamics. 

\section{Baryogenesis}

In this section we discuss the baryon number violation in the Randall-Sundrum braneworld context. 
The point is as follows. Since the bulk complex scalar field has the global U(1) symmetry in five 
dimensions, 
there is the current $J^M= i (\Phi \nabla^M \Phi^*-\Phi^* \nabla^M \Phi) $ 
satisfying the local conservation low $\nabla_M J^M=0$. However, this conservation does not 
mean the conservation of the projected current $J^\mu$ which is the observed quantity on the brane. 
Actually, we see $D_\mu J^\mu \sim -\partial_y J^y$. Depending on the value of 
$\partial_y J^y$, the projected current 
is not conserved in general. 

Let us define the the baryon number current by 
%
\begin{eqnarray}
J_\mu & := & i (\Phi \partial_\mu \Phi^*-\Phi^* \partial_\mu \Phi)|_{y=y_+} \nonumber \\
       & = &   2i (\eta_1 \partial_\mu \eta_2 -\eta_2 \partial_\mu \eta_1). 
\end{eqnarray}
%
Then the divergence of the current becomes
%
\begin{eqnarray}
{\cal D}_\mu J^\mu =2(\eta_1 {\cal D}^2 \eta_2 -\eta_2 {\cal D}^2 \eta_1). \label{div}
\end{eqnarray}
%
This is the equation on the positive tension brane. 
The right-hand side of this equation can be evaluated using the field equation. Since we are 
supposing that the spacetime dynamics is mainly governed by  
localized matters with $\stac{(\pm)}{T_{\mu\nu}}$ on the brane, we can see 
that Eq. (\ref{div}) becomes 
%
\begin{eqnarray}
{\cal D}_\mu J^\mu  & \simeq  & \ell^2 
{\cal D}_\mu \Biggl[\Biggl\lbrace \Bigl( \frac{-\frac{d_0}{\ell}}{1-a_0^2} +\frac{1}{2} \Bigr) 
\Bigl({}^{(4)}R^\mu_{\nu}(h)  -\frac{1}{3} \delta^\mu_\nu {}^{(4)}R \Bigr) \nonumber \\
& & +\frac{1}{4}(a_0^{-2}-1)
\Bigl({}^{(4)}G^{\mu}_{\nu}(h)-\kappa^2 \ell^{-1}\stac{(+)}{T^\mu_\nu} \nonumber \\
& & -\frac{\kappa^2}{2}({\cal D}^\mu \eta {\cal D}_\nu  \eta^* +{\cal D}^\mu \eta^* {\cal D}_\nu  \eta
-\delta^\mu_\nu |{\cal D}\eta|^2 ) \Bigr)
\Biggr\rbrace J^\nu 
\Biggr]\nonumber \\ \label{div2}
\end{eqnarray}
%
Therefore the current $J^\mu$ is not conserved in general. Now we consider the radiation dominated 
universe. 

From Eq. (\ref{div2}) we obtain the equation for the total charge $Q_B = \int d^3{\bf x} j^0$ 
%
\begin{eqnarray}
\dot Q_B \sim \ell^2 {}^{(4)} \dot R Q_B .
\end{eqnarray}
%
In the braneworld context, $\dot R$ is written as \cite{BWC}
%
\begin{eqnarray}
{}^{(4)} \dot R = -3(1+w)H \Biggl[ \frac{1-3w}{M_4^2}-\frac{(1+3w)\rho}{3M_5^6}  \Biggr] \rho, 
\end{eqnarray}
%
where $w=P/\rho $ and $H$ is the Hubble parameter. 
In the radiation dominated era, ${}^{(4)}\dot R \sim H T^{8}  M_5^{-6}$. 
Then it is easy to see that the rate of the baryon number violating process is given by 
%
\begin{eqnarray}
\Gamma_B \sim \ell^2 {}^{(4)} \dot R \sim \frac{\ell^2 H T^{8}}{M_5^6} \sim \frac{M_4^4 H T^{8}}{M_5^{12}}.
\end{eqnarray}
%
In the above we used $\ell \sim M_4^2/M_5^3$. Now we can estimate the decoupling 
temperature $T_D$ and then $T_D \sim M_5^{3/2}/M_4^{1/2}$. 
We suppose that 
$J^\mu$ is a current which produces net $B-L$. Then the finite baryon number density will be left 
after the electroweak sphaleron process occurs \cite{KRS}. 
The net baryon number density depends on the source of CP-violation. For example, 
we consider the CP violating interaction 
${\cal L}_{\rm int} \sim f M_*^{-2} \partial_\mu {}^{(4)} R J^\mu $ \cite{GB} 
which may be originated from the non-perturbative effect of the quantum gravity or 
the string theory (See Ref. \cite{BCKP} for another proposal.). 
$M_*$ is usually expected to be an corresponding scale to the bulk curvature, $\ell^{-1}$. 
$f$ is a factor which may be expected to be a small number. 
This interaction is the same as a CP violating interaction supposed in 
the spontaneous baryogenesis \cite{CK}. Actually, we can produce this interaction ${\cal L}_{\rm int}$ from 
the CP violating interaction term assumed in the usual spontaneous baryogenesis \cite{SK}. 
Interestingly, this interaction 
breaks the CPT spontaneously due to the dynamics of the expanding universe. Therefore, in this case, 
the out of 
thermal equilibrium is not necessary for the baryogenesis. 
 According to the same argument as Ref. \cite{SK}, the produced baryon to 
entropy ratio is \footnote{The baryon number escaped from the brane will back again to the same 
brane in Randall-Sundrum two brane system because the extra dimension is compactified. However, 
the point is the presence of the baryon number violating process on the brane. 
The details of this process is not important. C/CP violation is more important for the baryon number 
left in the present universe.} 
%
\begin{eqnarray}
n_B/s & \sim &f {}^{(4)} \dot R /M_*^2T|_{T=T_D} \nonumber \\
      & \sim & f (\ell M_* )^{-2}  H|_{T=T_D} T_D^{-1} \nonumber \\
      & = & f (M_5/M_*)^2(M_5/M_4)^{11/2}.
\end{eqnarray}
%
Since we should require $\ell^{-1} > {\rm TeV}$, a constraint on $M_5$ becomes 
$M_5 > 10^{41/3}{\rm GeV}$. The decoupling temperature becomes 
$T_D \sim 10^{11}(M_5/10^{41/3}{\rm GeV})^{3/2} {\rm GeV}$. 
If $M_* \sim \ell^{-1}$, then, we see 
$n_B/s \sim 10^{-10}(f/0.01)(M_5/10^{41/3}{\rm GeV})^{3/2}$. 
To obtain the reasonable value of 
$n_B/s$, we must set $M_5$ to be around $10^{14}{\rm GeV}$. This is consistent with 
the tabletop experimental lower bound $M_5 > 10^8 {\rm GeV}$ which is required on 
the positive tension brane.

\section{Summary}

In this paper we considered the baryogenesis in 
the Randall-Sundrum type model with the bulk complex scalar field. 
We found the baryon number violating process via the spacetime dynamics on the brane. 
Even if the scalar field do not have the potential, this mechanism can work. 
By assuming an appropriate source of CP violation, a 
reasonable net baryon number density can be 
obtained. However, the model that we considered here is only toy models. 
A study in a realistic and phenomenological model is needed.


\section*{Acknowledgements}

TU thanks A. Sugamoto for his continuous encouragement. SF thanks Y. Iwashita for 
useful discussions. The work of TS was supported by Grant-in-Aid for Scientific 
Research from Ministry of Education, Science, Sports and Culture of 
Japan(No.13135208, No.14740155 and No.14102004).

\end{document}